# Effect of initial condition of inflation on power and angular power spectra in finite slow-roll inflation

Shiro Hirai* and Tomoyuki Takami**

Department of Digital Games, Osaka Electro-Communication University

1130-70 Kiyotaki, Shijonawate, Osaka 575-0063, Japan

*Email: hirai@isc.osakac.ac.jp

**Email: takami@isc.osakac.ac.jp

## Abstract

The effect of the initial condition of inflation on the power spectra of scalar and tensor perturbations is estimated assuming a slow-roll inflation model. By defining a more general initial state in inflation, i.e., a squeezed state, particular properties of the power spectrum such as oscillation can be revealed. The behavior of the power spectrum is shown to exhibit a step-like variation with respect to finite inflation length in cases of both radiation- and scalar matter-dominated pre-inflation. The power spectrum is shown to oscillate in the radiation-dominated case. The effects of such a power spectrum on the TT and TE power spectra are examined for three typical slow-roll inflation models; a small-field model, a large field model, and a hybrid model, considering both pre-inflation models. It is found that the discrepancies between the three-year Wilkinson Microwave Anisotropy Probe (WMAP3) data and the $\Lambda$ CDM model, such as suppression of the spectrum at $l = 2$, may be explained to a certain extent by the finite length of inflation for inflation of close to 60 $e$-folds. The small-field inflation model with scalar matter-dominated pre-inflation provides the best fit to the WMAP3 data among the models considered. Relatively large changes in the angular TT power spectrum occur in response to small changes in inflation length in the radiation-dominated pre-inflation models, and half cases do not fit well with the observed data. This behavior is considered to be attributable to the oscillatory behavior of the power spectrum. The scalar matter-dominated case is thus preferred to the radiation-dominated case of pre-inflation, independent of the length of inflation. In both pre-inflation models, the angular TE power spectra at $l \leq 20$ are smaller in amplitude than that afforded by the $\Lambda$CDM model for inflation of close to 60 $e$-folds.

PACS number: 98.80.Cq

## 1. Introduction

Inflation [1] is an important concept in cosmology, and is supported by recent satellite-based measurements. However, consistent and natural models of inflation from the point of view of particle physics have yet to be established. The initial condition and length of inflation are of particular interest in the development of such a model, as the power spectrum is expected to be strongly affected by these two parameters in terms of a number of physical properties [2-4].

Information regarding the early universe can be extracted from the characteristics of the cosmic microwave background (CMB) and its anisotropy. Discrepancies between the data obtained by the Wilkinson Microwave Anisotropy Probe (WMAP) [5] and the widely accepted $\Lambda$ cold dark matter ($\Lambda$CDM) model have been studied by many researchers using a range of approaches, including trans-Planckian physics [4,6-8]. Such research has been conducted since the release of the first year of WMAP data, and many of the proposed concepts have survived even after the release of three-year WMAP data (WMAP3). However, WMAP3 data [9] provide more precise parameter values that deviate slightly from the best parameter values provided by first-year WMAP data for the $\Lambda$CDM model. For example, the reionization optical depth $\tau$ has changed from 0.17 to 0.089, and the scalar spectral index has decreased from 1.20 to 0.958. Although the $\Lambda$ CDM model comes close to explaining the WMAP3 data satisfactorily [9], there remains some inconsistency in the suppression of the spectrum at large angular scales ($l = 2$) and in the running of the spectral index. Importantly, the oscillatory behavior observed in the spectrum derived using first-year WMAP data has almost disappeared in the WMAP3 data set. The effect of this small discrepancy on inflation models and pre-inflation physics is thus an interesting problem.

The effect of the length of inflation and pre-inflation physics on the power spectrum and angular power spectrum of scalar and tensor perturbations was examined in previous studies considering power-law inflation [2-4]. In conventional analyses, the initial condition of inflation is typically assumed to be the Bunch-Davies state. However, a more general or squeezed state [10] could be applied in order to account for a wider variety of pre-inflation physics. A general formula for the power spectrum of scalar and tensor perturbations having any initial state in inflation was derived in a previous study as a familiar formulation multiplied by a factor indicating the contribution of the initial condition [2]. A subsequent paper [3] considered finite inflation models in which inflation began at a certain time, preceded by pre-inflation as a radiation-dominated or scalar matter-dominated period. In that study, calculations of the power spectrum were made for two matching conditions; one in which the gauge potential and its first $\eta$-derivative are continuous at the transition point (matching condition A), and another in which the transition occurs on a hyper-surface of constant energy (matching condition B), as proposed by Deruelle and Mukhanov [11]. That analysis revealed that when the length of inflation is finite, the power spectra can be expressed by a decreasing function at super-large scales, and that the power spectra oscillate from large to small scales in the radiation-dominated pre-inflation model and in both models under matching condition B [3]. The effect of such properties of the power spectrum on the angular temperature (TT) and temperature-polarization (TE) power spectra for finite power-law inflation has been investigated for these pre-inflation models using first-year WMAP data with respect to a scale factor defined by $a(\eta) = (-\eta)^p \propto t^q$, where $\eta$ is conformal time and $t$ is time [4]. Considering the two cases of pre-inflation and the two matching conditions, suppression of the TT spectrum at $l = 2,3$ was found to be explainable as due to the finite length of inflation for inflation of close to 60 $e$-folds in length with $q \geq 300$. Oscillatory behavior was observed in the radiation-dominated pre-inflation model and in both models under matching condition B. The angular TE power spectra thus derived differ somewhat from those afforded by the $\Lambda$CDM model at $l \leq 20$.

In the present study, this approach is applied to the analysis of slow-roll inflation models for comparison with WMAP3 data. By investigating the effects of the initial state of inflation on the power spectrum, some general properties of the power spectrum are derived. Three typical models of slow-roll inflation (small-field, large field, and hybrid models) are compared, and the effects of the length of inflation and pre-inflation physics (scalar matter- or radiation-dominated) on the angular TT and TE power spectra are investigated in detail. The derived TT and TE power spectra are compared with WMAP3 data and the $\Lambda$CDM model.

This paper is organized as follows. In Section 2, formulae for the power spectra of curvature perturbations and gravitational waves are derived for any initial condition in slow-roll inflation, and the properties of the formulae are derived. In Section 3, the power spectra for curvature perturbations and gravitational waves are derived using slow-roll inflation and the two pre-inflation models. In Section 4, using the derived formula, the angular TT and TE power spectra and the values of $\chi^2$ are calculated and compared with WMAP3 data. In Section 5, the results obtained in the present study are discussed at length.

## 2. Scalar and tenser perturbations

The formula for the power spectrum of curvature perturbations in inflation is derived here for any initial condition by applying a commonly used method [12]. Although this formula was originally derived in Refs. [13, 2], as a critical result it is derived explicitly in the present study for the case of slow-roll inflation, and a new property of the power spectrum is introduced. The background spectrum considered is a spatially flat Friedman-Robertson-Walker (FRW) universe described by metric perturbations. The line element for the background and perturbations is generally expressed as [14]

$$ds^2 = a^2(\eta)\{(1+2A)d\eta^2 - 2\partial_i B dx^i d\eta - [(1-2\Psi)\delta_{ij} + 2\partial_i\partial_j E + h_{ij}]dx^i dx^j\}, \quad (1)$$

where $\eta$ is the conformal time, the functions $A$, $B$, $\Psi$, and $E$ represent the scalar perturbations, and $h_{ij}$ represents tensor perturbations. The density perturbation in terms of the intrinsic curvature perturbation of comoving hypersurfaces is given by $\Re = -\Psi - (H/\dot{\phi})\delta\phi$, where $\phi$ is the inflaton field, $\delta\phi$ is the fluctuation of the inflaton field, $H$ is the Hubble expansion parameter, and $\Re$ is the curvature perturbation. Overdots represent derivatives with respect to time $t$, and the prime represents the derivative with respect to the conformal time $\eta$. Introducing the gauge-invariant potential $u \equiv a(\eta)(\delta\phi + (\dot{\phi}/H)\Psi)$ allows the action for scalar perturbations to be written as [15]

$$S = \frac{1}{2}\int d\eta d^3x \{(\frac{\partial u}{\partial \eta})^2 - c_s^2 (\nabla u)^2 + \frac{Z''}{Z}u^2 \}, \quad (2)$$

where $c_s$ is the velocity of sound, $Z = a\dot{\phi}/H$, and $u = -Z\Re$. Next, the tensor perturbations are considered, where $h_{ij}$ represents gravitational waves. Under the transverse traceless gauge, the action of gravitational waves in the linear approximation is given by [14]

$$S = \frac{1}{2} \int \mathrm{d}^4 x \, \{ (\frac{\partial h}{\partial \eta})^2 - (\nabla h)^2 + \frac{a''}{a} \, h^2 \}, \tag{3}$$

where $h$ is the transverse traceless part of the deviation of $h_{ij}$ and represents the two independent polarization states of the wave ($h_+, h_\times$). The fields $u(\eta, \boldsymbol{x})$ and $h(\eta, \boldsymbol{x})$ are expressed using annihilation and creation operators as follows.

$$u(\eta, \, \boldsymbol{x}) = \frac{1}{(2\pi)^{3/2}} \int \mathrm{d}^3 k \, \{ u_k(\eta) \, \boldsymbol{a_k} + u_k *(\eta) \, \boldsymbol{a_{-k}}^\dagger \} \, e^{-i \, kx} \,, \tag{4}$$

$$h(\eta, \, \boldsymbol{x}) = \frac{1}{(2\pi)^{3/2} a(\eta)} \int \mathrm{d}^3 k \, \{ v_k(\eta) \, \boldsymbol{a_k} + v_k *(\eta) \, \boldsymbol{a_{-k}}^\dagger \} \, e^{-i \, kx}. \tag{5}$$

The field equation for $u_k(\eta)$ is derived as

$$\frac{\mathrm{d}^2 u_k}{\mathrm{d}\eta^2} + (c_s^2 \, k^2 - \frac{1}{Z} \frac{\mathrm{d}^2 Z}{\mathrm{d}\eta^2}) \, u_k = 0 \,. \tag{6}$$

The field equation for $v_k(\eta)$ becomes eq. (6) with $c_s^2 = 1$ and $Z = a(\eta)$. The solution to $u_k$ and $v_k$ satisfy the normalization condition $u_k \mathrm{d}u_k^* / \mathrm{d}\eta - u_k^* \mathrm{d}u_k / \mathrm{d}\eta = i$. For slow-roll inflation, the following parameters are employed [12,16].

$$\varepsilon = 3 \frac{\dot{\phi}^2}{2} (\frac{\dot{\phi}^2}{2} + V)^{-1} = \frac{m^2}{4\pi} \left( \frac{H'(\phi)}{H(\phi)} \right)^2 \,, \tag{7}$$

$$\delta = \frac{m^2}{4\pi} \frac{H''(\phi)}{H(\phi)} \,, \tag{8}$$

$$\xi = \frac{m^4}{16\pi^2} \frac{H'(\phi) H'''(\phi)}{(H(\phi))^2} = \frac{\dot{\varepsilon} - \dot{\delta}}{H} \,. \tag{9}$$

The quantity $V(\phi)$ is the inflation potential, and $m$ is the Plank mass. Other slow-roll parameters ($\varepsilon_V, \eta_V, \xi_V$) can be written in terms of the slow-roll parameters $\varepsilon$, $\delta$, and $\xi$ to first order in slow roll: $\varepsilon = \varepsilon_V$, $\delta = \eta_V - \varepsilon_V$, and $\xi = \xi_V - 3\varepsilon_V \eta_V + 3\varepsilon_V^2$, where

$\varepsilon_V = m^2/16\pi(V'/V)^2$ , $\eta_V = m^2/8\pi(V''/V)$ , and $\xi_V = m^4/64\pi^2(V'V'''/V^2)$ . Using the slow-roll parameters, $(\mathrm{d}^2 Z/\mathrm{d}\eta^2)/Z$ is written exactly as

$$\frac{1}{Z}\frac{\mathrm{d}^2 Z}{\mathrm{d}\eta^2} = 2a^2 H^2 \,(1 + \varepsilon - \frac{3}{2}\delta + \varepsilon^2 - 2\varepsilon\delta + \frac{\delta^2}{2} + \frac{\xi}{2})\,, \tag{10}$$

and the scale factor is written as $a(\eta) = -((1-\varepsilon)\eta H)^{-1}$. Here, it is assumed that the slow-roll parameters satisfy $\varepsilon < 1, \delta < 1$, and $\xi < 1$. As only the terms of leading order of $\varepsilon$ and $\delta$ are adopted, it may be considered that $\varepsilon$ and $\delta$ are constant, allowing the scale factor to be written as $a(\eta) \approx (-\eta)^{-1-\varepsilon}$ [16]. Equation (6) can then be rewritten as

$$\frac{\mathrm{d}^2 u_k}{\mathrm{d}\eta^2} + (k^2 - \frac{2 + 6\varepsilon - 3\delta}{\eta^2})u_k = 0\,, \tag{11}$$

where $c_s{}^2 = 1$ in the scalar field case. The solution for eq. (11) is then written as

$$f_k^I(\eta) = i\frac{\sqrt{\pi}}{2}e^{-ip\pi/2}(-\eta)^{1/2}\,H_{-p+1/2}^{(1)}(-k\eta)\,, \tag{12}$$

where $p = -1 - 2\varepsilon + \delta$ , and $H_{-p+1/2}^{(1)}$ is the Hankel function of the first kind with order $-p+1/2$ . In the gravitational case, $v_k(\eta)$ becomes eq. (12) with $p = -1 - \varepsilon$ . As a general initial condition, the mode functions $u_k(\eta)$ and $v_k(\eta)$ are assumed to be

$$u_k(\eta) = c_1\,f_k^I(\eta) + c_2\,f_k^{I*}(\eta)\,, \tag{13}$$

$$v_k(\eta) = c_{g1}\,f_k^I(\eta) + c_{g2}\,f_k^{I*}(\eta)\,, \tag{14}$$

where the coefficients $c_1$ and $c_2$ ( $c_{g1}$ and $c_{g2}$ ) obey the relation $|c_1|^2 - |c_2|^2 = 1$ ( $|c_{g1}|^2 - |c_{g2}|^2 = 1$ ). The important point here is that the coefficients $c_1$ and $c_2$ ( $c_{g1}$ and $c_{g2}$ ) do not change during inflation. In ordinary cases, the field $u_k(\eta)$ is considered to be in the

Bunch-Davies state, i.e., $c_1 = 1$ and $c_2 = 0$, because as $\eta \to -\infty$, the field $u_k(\eta)$ must approach plane waves ($e^{-ik\eta}/\sqrt{2k}$).

Next, the power spectra of the scalar $P_{\Re}$ and tensor $P_g$ are defined as follows [12].

$$< \Re_k(\eta), \Re_k *(\eta) >= \frac{2\pi^2}{k^3} P_{\Re} \, \delta^3 (\boldsymbol{k\text{-}l}), \tag{15}$$

$$< v_k(\eta) v_l *(\eta) >= \frac{\pi m^2 a^2}{16 k^3} P_g \, \delta^3 (\boldsymbol{k\text{-}l}), \tag{16}$$

where $\Re_k(\eta)$ is the Fourier series of the curvature perturbation $\Re$. The power spectra $P_{\Re}^{1/2}$ and $P_g^{1/2}$ are then written as follows [12].

$$P_{\Re}^{1/2} = \sqrt{\frac{k^3}{2\pi^2}} \left| \frac{u_k}{Z} \right|, \tag{17}$$

$$P_g^{1/2} = \frac{4\sqrt{k^3}}{m\sqrt{\pi} a} |v_k|. \tag{18}$$

Using the approximation of the Hankel function, the power spectra of the leading and next-leading corrections of $-k\eta$ in the case of the squeezed initial states (13) and (14) can be written as

$$P_{\Re}^{1/2} = (2^{-p} (-p)^p \frac{\Gamma(-p+1/2)}{\Gamma(3/2)} \frac{1}{m^2} \frac{H^2}{|H'|}) |_{k=aH} \; (1 - \frac{(-k\eta)^2}{2(1+2p)}) |c_1 \, e^{-ip\pi/2} + c_2 \, e^{ip\pi/2} |$$

$$\cong (\frac{H^2}{2\pi\dot{\phi}}) |_{k=aH} |c_1 \, e^{-ip\pi/2} + c_2 \, e^{ip\pi/2} |, \tag{19}$$

$$P_g^{1/2} = (\frac{2^{-q+1}}{\sqrt{\pi}} (-q)^q \frac{\Gamma(-q+1/2)}{\Gamma(3/2)} \frac{H}{m}) |_{k=aH} \; (1 - \frac{(-k\eta)^2}{2(1+2q)}) |c_{g1} \, e^{-iq\pi/2} + c_{g2} \, e^{iq\pi/2} |,$$

$$\tag{20}$$

where $p = -1 - 2\varepsilon + \delta$, $q = -1 - \varepsilon$, and $\Gamma(-p + 1/2)$ represents the Gamma function. $P_g^{1/2}$ is multiplied by a factor of $\sqrt{2}$ to account for the two polarization states. These formulae differ slightly from Hwang's formula [13] due to the introduction of the term $e^{-ip\pi/2}$ into eq. (12), as required such that in the limit $\eta \to -\infty$, the field $u_k(\eta)$ must approach plane waves. The quantities $C(k)$ and $C_g(k)$ are defined as

$$C(k) = c_1 \, e^{-ip\pi/2} + c_2 \, e^{ip\pi/2}, \tag{21}$$

$$C_g(k) = c_{g1} \, e^{-iq\pi/2} + c_{g2} \, e^{iq\pi/2}. \tag{22}$$

The general properties of the quantity $|C(k)|$ are now investigated. Using the squeezing parameters $R$, $\varphi$ and $\alpha$ the coefficients $c_1$ and $c_2$ can be written as $c_1 = e^{i\alpha} \cosh R$ and $c_2 = e^{i\alpha} e^{i\varphi} \sinh R$. The quantity $|C(k)|^2$ is written in terms of $R$ and $\varphi$ as

$$|C(k)|^2 = \cosh 2R + \sinh 2R \cos(p\pi + \varphi). \tag{23}$$

In the tensor case, changing $p$ for $q$, a similar relation is derived. When $R = 0$, $|C(k)|^2$ becomes 1, but in many other cases $|C(k)|^2 > 1$. If inflation starts with a squeezed state owing to pre-inflation physics, it appears that $|C(k)|^2 > 1$, and in ordinary cases it is inferred that $R$ and $\varphi$ are dependent on $k$. Thus, $|C(k)|^2$ may oscillate in many cases, since $\cosh R$ and $\sinh R$ are of the same order except when $R$ is near 0. This oscillatory property of the power spectrum is demonstrated in the next section.

## 3. Calculation of power spectrum

In Section 2, the initial states of the scalar and tensor perturbations were derived as general (squeezed) states (eqs. (13) and (14)), rather than assuming the conventional Bunch-Davies vacuum for the initial state of inflation. To illustrate some of the general properties of the power spectrum, the case of slow-roll inflation of finite length is examined with respect to the effect of the length of inflation and the pre-inflation physics on the power spectrum. The pre-inflation model is considered to consist simply of a radiation-dominated period or a scalar matter-dominated period. A simple cosmological model is assumed, as defined by

$$\text{Pre-inflation:} \qquad a_P(\eta) = b_1 (-\eta - \eta_j)^r ,$$

$$\text{Inflation:} \qquad a_1(\eta) = b_2 (-\eta)^{-1-\varepsilon} , \qquad (24a)$$

where

$$\eta_j = -(\frac{r}{1+\varepsilon}+1)\eta_2 , \ b_1 = (\frac{-1-\varepsilon}{r})^r (-\eta_2)^{-1-\varepsilon-r} b_2 . \qquad (24b)$$

The scale factor $a_1(\eta)$ represents slow-roll inflation. Inflation is assumed to begin at $\eta = \eta_2$. In pre-inflation, for the case of $r = 1$, the scale factor $a_P(\eta)$ indicates that pre-inflation is a radiation-dominated period, whereas for the case of $r = 2$, the scale factor $a_P(\eta)$ indicates a scalar matter-dominated period.

The curvature and tensor perturbations in the scalar matter and radiation cases are calculated as follows. In the case of curvature perturbations, the field equation $u_k$ can be written as eq. (6) with the value $c_s{}^2 = 1/3$ in the case of radiation and $c_s{}^2 = 1$ in the case of scalar matter, and $Z = a_P(\eta)[2(\boldsymbol{H}^2 - \boldsymbol{H}')/3]^{1/2}/(c_s \boldsymbol{H})$ [15,18], where $\boldsymbol{H} = a_P{}'/a_P$ in both the pre-inflation models. The solution to eq. (6) is then written as $f_k^S(\eta) = (1 - i/(k(\eta + \eta_j))) \exp[-ik((\eta + \eta_j)]/\sqrt{2k}$ for the scalar matter-dominated case, and $f_k^R(\eta) = 3^{1/4} \exp[-ik(\eta + \eta_i)/\sqrt{3}]/\sqrt{2k}$ for the radiation-dominated case. For gravitational waves, the solution to eq. (6) is written as $f_k^{GS}(\eta) = (1 - i/(k(\eta + \eta_j))) \exp[-ik((\eta + \eta_j)]/\sqrt{2k}$ for the scalar matter-dominated case, and $f_k^{GR}(\eta) = \exp[-ik(\eta + \eta_j)]/\sqrt{2k}$ for the radiation-dominated case. In previous papers [2-4], the Bunch-Davies state was assumed for pre-inflation. However, to allow for a more general physical history of the universe, the pre-inflation state is represented in the present study using the squeezing parameters $X$ and $\theta$. As concrete values cannot be determined at present, a general feature is derived from such squeeze states.

In the scalar matter-dominated case of pre-inflation, the general states of the scalar and tensor perturbations are written respectively as follows.

$$u_k(\eta) = \cosh X \; f_k^S(\eta) + \sinh X \; e^{i\theta} \; f_k^S *(\eta), \tag{25}$$

$$v_k(\eta) = \cosh X' \; f_k^{GS}(\eta) + \sinh X' \; e^{i\theta'} \; f_k^{GS} *(\eta). \tag{26}$$

The coefficients $c_1, c_2, c_{g1}$, and $c_{g2}$ are fixed using the matching condition in which the mode function and first $\eta$-derivative of the mode function are continuous at the transition time $\eta = \eta_2$ ($\eta_2$ is the beginning of inflation). The coefficients $c_1, c_2, c_{g1}$, and $c_{g2}$ can then be calculated analytically, and $C(k)$ and $C_g(k)$ can be derived from eqs. (21) and (22). The quantity $|C(k)|^2$ can be calculated analytically by

$$|C(k)|^2 = A_1 \cosh 2X - \left( A_2 \cos\left(\frac{4z}{q} - \theta\right) + A_3 \sin\left(\frac{4z}{q} - \theta\right) \right) \sinh 2X, \tag{27}$$

where

$$A_1 = \frac{\pi}{128z^3} \left\{ 16z^2 \left( (2+q-2p)B_1 - 2zB_2 \right)^2 + \left( -2\left(q^2 - 2q(-1+p) - 4z^2\right)B_1 + 4qzB_2 \right)^2 \right\},$$

$$A_2 = \frac{\pi}{128z^3} \left\{ -16z^2 \left( (2+q-2p)B_1 - 2zB_2 \right)^2 + \left( -2\left(q^2 - 2q(-1+p) - 4z^2\right)B_1 + 4qzB_2 \right)^2 \right\},$$

$$A_3 = \frac{\pi}{8z^2} \left\{ \left( (2+q-2p)B_1 - 2zB_2 \right)\left( \left(q^2 - 2q(-1+p) - 4z^2\right)B_1 - 2qzB_2 \right) \right\},$$

$B_1 = BesselJ[1/2 - p, z]$, $B_2 = BesselJ[3/2 - p, z]$, $z = -k\eta_2$.

For $z \to 0$, the leading term is written as

$$|C(k)|^2 \approx \frac{4^{-3+p} q^2 \pi (2+q-2p)^2 z^{-2(1+p)}}{\Gamma(3/2-p)^2} \left\{ \cosh 2X - \cos\left(\frac{4z}{q} - \theta\right) \sinh 2X \right\}. \quad (28)$$

For $z \to \infty$, the leading term is written as

$$|C(k)|^2 \approx \cosh 2X + \cos\left(p\pi + (2-4/q)z + \theta\right) \sinh 2X, \quad (29)$$

where $p = -1 - 2\varepsilon + \delta$, and $q = -1 - \varepsilon$. Equation (29) has an interesting property. When $X = 0$, $|C(k)|^2$ becomes 1 in the case of $z > 1$. However, when the state of scalar matter in pre-inflation is not the Bunch-Davies state (i.e., squeezing occurs), $|C(k)|^2 > 1$ and oscillatory properties emerge because $\cosh 2X$ and $\sinh 2X$ are of the same order ($X$ is not near 0). A similar property was discussed in Section 2. An important point is that even when the squeezing value is small (i.e., $\sinh 2X \cong 1$), the oscillatory property remains from large to small scales. This property is clearly different from the case of $X = 0$, and is likely to affect the angular power spectrum. As shown later, the radiation-dominated pre-inflation model is a typical example. The quantity $C_g(k)$ can be derived for the tensor perturbation case by substituting $p$, $X$, and $\theta$ for $q$, $X'$, and $\theta'$ in eqs. (27)–(29). Similar behavior to the scalar perturbations is thus derived.

In the radiation-dominated case, $|C(k)|^2$ can be calculated analytically by

$$|C(k)|^2 = A_4 \cosh 2X + \left(A_5 \cos\left(-\frac{2z}{\sqrt{3}q} + \theta\right) + A_6 \sin\left(-\frac{2z}{\sqrt{3}q} + \theta\right)\right) \sinh 2X, \quad (30)$$

where

$$A_4 = \frac{\pi}{2\sqrt{3}z}(6(p-1)zB_1B_2 + 3z^2B_2^2 + (3-6p+3p^2+z^2)B_1^2),$$

$$A_5 = \frac{\pi}{2\sqrt{3}z}(6(p-1)zB_1B_2 + 3z^2B_2^2 + (3-6p+3p^2-z^2)B_1^2),$$

$$A_6 = \pi B_1((p-1)B_1 + zB_2),$$

$$B_1 = BesselJ\left[1/2 - p, z\right], \; B_2 = BesselJ\left[3/2 - p, z\right], \; z = -k\eta_2 \,.$$

For $z \to 0$, the leading term is written as

$$|C(k)|^2 \approx \frac{\sqrt{3}\pi 4^{-1+p}(p-1)^2 z^{-2p}}{\Gamma(3/2-p)^2}\left\{\cosh 2X + \cos\left(-\frac{2z}{\sqrt{3}q} + \theta\right)\sinh 2X\right\}. \tag{31}$$

For $z \to \infty$, the leading term is written as

$$|C(k)|^2 \approx \frac{2 + \cos(p\pi + 2z)}{\sqrt{3}}\cosh 2X + \frac{1 + 2\cos(p\pi + 2z)}{\sqrt{3}}\cos(-\frac{2z}{\sqrt{3}q} + \theta)\sinh 2X. \tag{32}$$

The difference from the scalar matter-dominated case is that even if $X = 0$, the power spectrum oscillates at $z > 1$ (from eq. (32)), but the amplitude is of order 1. Even if inflation is very long, this oscillation does not disappear. The effect on the angular power spectrum is shown later. In the case of $X \neq 0$, the amplitude may increase and complex oscillation may occur. As this case is highly model-dependent, it is not discussed any further. In the tensor case, $|C_g(k)|^2$ can be derived in a similar manner to eqs. (27)-(29), as shown in the appendix.

The properties of the derived power spectrum are examined here in detail for the two pre-inflation models. The power spectrum of scalar perturbations in the scalar matter-dominated pre-inflation case with $X = 0$ exhibits the following properties. As $z \to \infty$, the power spectrum approaches 1 (from eq. (29)), while as $z \to 0$, $|C(k)|^2 \approx z^{4\varepsilon-2\delta}$ (eq. (28)). When $2\varepsilon > \delta$, $|C(k)|^2$ becomes zero as $z \to 0$, whereas when $2\varepsilon < \delta$, $|C(k)|^2$ becomes infinity as $z \to 0$. Within the parameter ranges of $0 \leq \varepsilon \leq 1$ and $\varepsilon \leq \delta \leq 1$, $|C(k)|^2$ is close to 1 at $z > 2.0$, and as $z \to 0$, $|C(k)|^2$ initially decreases then gradually increases. This is an example of a finite hybrid inflation model. In the case of $2\varepsilon = \delta$, the $z$-dependence is lost, and $|C(k)|^2$ becomes $0.0625(1-\varepsilon/3)^2(1+\varepsilon)^2$ as $z \to 0$. The coefficient of eq. (28), $4^{-4+\delta-2\varepsilon}\pi(1+\varepsilon)^2(3-2\delta+3\varepsilon)^2/\Gamma(5/2-\delta+2\varepsilon)^2$, is less than 0.1 at $0 \leq \varepsilon \leq 1$ and $0 \leq \delta \leq 1$.

This small value becomes important as the power spectrum decreases in amplitude by a factor of 10 from $z = 2$ to $z = 0$. This behavior is considered be one of the physical origins for the cut-off of the power spectrum [8]. To illustrate the contribution of finite inflation and the domination of scalar matter to pre-inflation physics, the power spectra for three inflation models are plotted in Fig. 1. The three models are a small-field model with $\varepsilon = 0.026$ and $\delta = -0.0375$, a $\phi^2$ model with $\varepsilon = 0.00833$ and $\delta = 0$, and a hybrid model with $\varepsilon = 0.002$ and $\delta = 0.0198$ as a typical slow-roll inflation model. The spectra for the small-field and $\phi^2$ models are very similar, whereas the power spectrum in the hybrid case becomes larger as $z \to 0$, representing the contribution of $4\varepsilon - 2\delta < 0$. The power spectrum shape afforded by the two former models, as used in many previous studies [7,8], results in lower values of the angular power spectrum at $l = 2,3$ compared to the $\Lambda$CDM model. This shape of the power spectrum is similar to that for gravitational waves in both scalar matter- and radiation-dominated pre-inflation models.

In the radiation-dominated pre-inflation case with $X = 0$, $|C(k)|^2$ oscillates between $1/\sqrt{3}$ and $\sqrt{3}$ at $z > 2.0$, and becomes zero as $z \to 0$ within the parameter ranges of $0 \leq \varepsilon \leq 1$ and $0 \leq \delta \leq 1$. In this case, the coefficient of eq. (31) is constrained by $0.0025 \leq \sqrt{3}\pi 4^{-2+\delta-2\varepsilon}(\delta - 2\varepsilon - 2)^2 / \Gamma(5/2 - \delta + 2\varepsilon)^2 \leq 1.73$. When $\varepsilon$ and $\delta$ are close to zero, this coefficient is close to 1. As in the scalar matter-dominated case, $|C(k)|^2$ exhibits a step-like variation in the range of $0 \leq z \leq 2$. In this case, the contribution of $z^{2+4\varepsilon-2\delta}$ appears to be essential. As there is little difference among the three inflation models in the radiation-dominated case, the power spectrum for the small-field model is plotted in Fig. 2 as a typical example of the slow-roll inflation model.

In the above discussion, the case of $X = 0$ was considered. However, it can be presumed from the form of eqs. (28)-(32) that the properties of other cases ($X \neq 0$) will be similar to those obtained for the $X = 0$ case. Similar properties of the power spectrum have been reported in previous papers considering finite power-law inflation [2,3].

The variable $z = -k\eta_2$ should at this point be examined in more detail. A value of $z = 1$ corresponds to the case that inflation starts at $\eta_2$, denoting the time at which the perturbations of the present Hubble horizon size exceed the Hubble radius in inflation. Figures 1 and 2 show plots for an inflation of 60 $e$-folds in length. If inflation continues for $a$ times longer, the perturbation of the current Hubble horizon size is obtained by $z \approx a$. In the case of radiation-dominated pre-inflation, if a longer inflation ($a > 1$) exists, $|C(k)|^2$ behaves according to eq. (32) from the super horizon scales to small scales, oscillating around $2/\sqrt{3}$ (see Fig. 2). It is important to note that this oscillation persists even for very a long inflationary period. It can be inferred from the discussions above that such oscillatory behavior will arise in many cases in which the state of the scalar perturbations is written according to eq. (25). For example, in the case when the inflation transition is required to occur on a hyper-surface of constant energy [11], strong oscillation has also been found to occur [3].

The ratio of gravitational waves to curvature perturbations in the power spectrum is given by

$$R(k) = \frac{P_g}{P_{\Re}} \approx 16\varepsilon \frac{|C_g(k)|^2}{|C(k)|^2},$$
(33)

where the term $16\varepsilon$ represents the contribution of slow-roll inflation. The quantity $|C_g(k)|^2 / |C(k)|^2$ ($= R_c$) is dependent on the pre-inflation model and the values of $\varepsilon$ and $\delta$. In the scalar matter-dominated pre-inflation model, $R_c$ oscillates toward 1 as $z \to \infty$. Since $R_c \approx z^{-2\varepsilon+2\delta}$, as $z \to 0$, $R_c$ becomes infinite in the small-field and $\phi^2$ models, and zero in the hybrid model. However, it appears that this divergent behavior of $R_c$ as $z \to 0$ has little effect on the angular power spectrum since the dominant contribution of the calculation of the $l$ th angular power spectrum $C_l$ appears at $z \approx la/3.3$ [19]. The scalar-tensor ratio $R(k)$ at $l = 2$ is required for calculation of the angular power spectrum. At $z \approx 1$, $R_c$ is close to 1 in the scalar matter-dominated pre-inflation case. Therefore, $R_c$ is fixed at 1 in the calculation of the angular power spectrum.

In the radiation-dominated pre-inflation case, $R_c$ oscillates in a manner comparable to $\sqrt{3}/(2-\cos\theta)$ (i.e., $1/\sqrt{3} \leq R_c \leq \sqrt{3}$ ) as $z \to \infty$, while the behavior as $z \to 0$ is similar to that exhibited by the scalar matter-dominated model. At $z \approx 1$, $R_c \approx 0.6$ in the case of the small-field model. Therefore, $R_c(k)$ is fixed at 0.6 in the calculation of the angular power spectrum. It should be noted, however, that the behavior of the angular power spectrum with $R_c(k) = 1$ is not appreciably different from this case.

## 4. Angular power spectra

Calculation of the power spectra for scalar and tensor perturbations under the two pre-inflation models for the case of slow-roll inflation reveals two interesting properties. First, when the length of inflation is finite, the power spectrum can be expressed by a step-like function in the region of $0 \leq z \leq 2$. This behavior is seen in all of the cases considered in the present study. Second, the scalar matter-dominated pre-inflation models afford power spectra that reach an amplitude of 1 as $z \to \infty$, while the power spectrum for the radiation-dominated case oscillates from large to small scales (Figs. 1 and 2).

The effect of this behavior on the angular power spectrum is investigated by comparison of the calculated angular power spectra with three-year WMAP data. The angular TT and TE power spectra $C_l$ and $C_l^{\text{TE}}$ can be written as [20]

$$C_l = 4\pi \int T_\Theta^2(k,l) \ P_\Re(k) \frac{\mathrm{d}k}{k} , \tag{34}$$

$$C_l^{\text{TE}} = 4\pi \int T_\Theta(k,l) T_{\text{E}}(k,l) \ P_\Re(k) \frac{\mathrm{d}k}{k} , \tag{35}$$

where $T_\Theta(k,l)$ and $T_{\text{E}}(k,l)$ are transfer functions. The angular TT and TE power spectra were computed in the present study using a modified CMBFAST code [21] assuming a flat universe with the following parameter values: baryon density $\Omega_b = 0.0412$, dark energy density $\Omega_\Lambda = 0.768$, present-day expansion rate $H_0 = 73.4$ km s$^{-1}$ Mpc$^{-1}$, and reionization optical depth $\tau = 0.091$ (from WMAP3 data). The present treatment considers the effect of the length of inflation $a$, where $a = 1$ indicates that inflation starts at the time when the perturbations of the current Hubble horizon size exceed the Hubble radius in inflation (i.e., inflation of close to 60 $e$-folds). From eqs. (27) and (30), $|C(k)|^2$ and $|C_g(k)|^2$ can be derived analytically using the Bessel function. In the present calculations of the angular power spectra, the case of $X = 0$ is considered and the expansions of the Bessel function in $|C(k)|^2$ and $|C_g(k)|^2$ at $z = 0$ and $z = \infty$ are used. The angular TT power spectra are normalized with respect to 11 data points in the WMAP3 data from $l = 65$ to $l = 210$ so as to average the small changes in $l$ due to the models and the contribution of oscillation. The same values are used in the analysis of the angular TE spectrum. In the likelihood analysis, the values of $\chi^2$ are calculated for the angular TT and TE power spectra against data points 987 and 426 in the WMAP3 data using the WMAP Likelihood Code [22].

Three slow-roll inflation models are considered as discussed above; a small-field model, a large field model ($\phi^2$ model), and a hybrid inflation model. The potential terms are written as $V = \Lambda^4(1-(\phi/\mu)^2)$ for the small-field model, $V(\phi) = \Lambda^4(\phi/\mu)^2$ for the large field model, and $V = \Lambda^4(1+(\phi/\mu)^2)$ for the hybrid model. The effect of the length of inflation on the angular power spectrum is examined using these models with the relevant values of $\varepsilon$ and $\delta$.

(1) *Scalar matter-dominated pre-inflation*

In the scalar matter-dominated case, the length of inflation is critical to suppression of the angular TT power spectrum of the CMB with $l = 2$ in all three slow-roll inflation models. In the small-field model (Fig. 3), with $n_s = 0.961275$, $\mathrm{d}\,n_s/\ln k = -\,0.00002454$, $n_t = -0.0052369$, $\mathrm{d}\,n_t/\mathrm{d}\ln k = -0.00001083$, and $r_2 = 0.03613$ (derived from the potential with $\mu = 15m$), the CMB power spectrum is suppressed compared to that of the $\Lambda\,\text{CDM}$ model. The suppression decreases from 51% at $a = 0.5$ to 12% at $a = 2.0$. The $l = 2$ spectrum at $a = 2.0$ fits the WMAP3 data within the limit of error, while the $\Lambda\,\text{CDM}$ model is most consistent with the spectra with $a > 3$. As shown in Fig. 3, the spectrum produced in the case of $a = 10.0$ cannot be differentiated from that given by the $\Lambda\,\text{CDM}$ model. The best values of $\chi^2$ are obtained for $a = 1.2$: $\chi^2\,(\text{TT}) = 1056.47$, $\chi^2\,(\text{TE}) = 417.50$, $\chi^2\,(\text{total}) = 1473.97$. However, the value of $\chi^2$ varies little in the range of $0.8 \le a \le 10.0$: for example, at $a = 10.0$, $\chi^2\,(\text{TT}) = 1056.51$, $\chi^2\,(\text{TE}) = 417.49$, $\chi^2\,(\text{total}) = 1474.00$. The best values of $\chi^2$ given by the $\Lambda\,\text{CDM}$ model are $\chi^2\,(\text{TT}) = 1056.68$, $\chi^2\,(\text{TE}) = 418.34$, and $\chi^2\,(\text{total}) = 1475.02$, which are obtained with parameters of $\Omega_b = 0.0416$, $\Omega_\Lambda = 0.7617$, $H_0 = 73.2$ km s$^{-1}$ Mpc$^{-1}$, and $\tau = 0.089$ (from WMAP3 data).

In the $\phi^2$ model, the length of inflation is also critical. Suppression of the angular power spectrum is obtained with $a < 2$. As $a$ becomes smaller, $C_l^{\mathrm{TT}}$ becomes larger than that obtained by the $\Lambda$CDM model in the range $20 \leq l \leq 50$, causing $\chi^2$ to becomes worse.

In the hybrid inflation model with typical values of $p = 2$ and $n_s = 1.04$, no reasonable agreement with the WMAP3 data could be obtained for lengths of inflation from $a = 1$ to $a = 10$. A similar result has been reported previously [23]. It appears that the divergent behavior of the power spectrum at $z \rightarrow 0$ has little effect on the power spectra for small $l$ in the present case.

The results above suggest that the small-field inflation model provides better agreement with the WMAP3 data than either the $\phi^2$ or hybrid inflation models.

(2) *Radiation-dominated pre-inflation*

The power spectrum afforded by a radiation-dominated pre-inflation model exhibits oscillation from large to small scales (see Fig. 2), and the influence of this oscillation on the angular power spectrum is very interesting. The angular TT power spectrum displays an oscillation (at $l \geq 5$), which may explain the small oscillation seen in the angular TT power spectrum of the WMAP3 data (should it be confirmed to exist). The change in spectral shape, including the shape of the first peak, indicates a strong dependence on the value of $a$. The change in the shape of the TT power spectrum exhibits periodicity with respect to $a$, having a cycle of 7.3 in the small-field case. This phenomenon occurs out to $a \approx 200$, yet is suppressed at low $l$ in the range of $a < 2$ as in the scalar matter-dominated case. Figure 4 shows the TT power spectrum for $a = 0.75$, which results in suppression of the spectrum at $l = 2$ and better $\chi^2$. In this case, $\chi^2(\mathrm{TT}) = 1095.35$, $\chi^2(\mathrm{TE}) = 417.22$, and $\chi^2(\mathrm{total}) = 1512.57$. The fit is therefore relatively poor. The shape of the angular TT power spectrum thus changes substantially with a small change in the length of inflation in the radiation-dominated model, and half-*a* cases afford poor $\chi^2$ values. Such properties can be attributed to the oscillatory behavior of the power spectrum. Although the radiation-dominated case should not be completely ruled out because of this sensitive *a* dependence, the scalar matter-dominated model is preferred to the radiation-dominated pre-inflation model.

(3) *Angular TE power spectrum*

The angular TE power spectrum for $l \geq 20$ is similar to that given by the $\Lambda$ CDM model in almost all cases considered here. However, differences with respect to the length of inflation emerge: at $a < 3$ the values of $C_l^{TE}$ are smaller than in the $\Lambda$ CDM model at $l \leq 20$, and if $C_l^{TT}$ exhibits large oscillation, $C_l^{TE}$ undergoes small oscillation in the radiation-dominated case. The angular TE power spectrum for $l \leq 20$ assuming a small-field model and scalar matter-dominated pre-inflation is plotted in Figure 5 for various values of *a* as an example of these results.

## 5. Discussion and summary

The effects of the initial condition of inflation and the length of inflation on the power and angular power spectra were examined in the present study for the case of slow-roll inflation. Whereas the Bunch-Davies vacuum is widely assumed for the initial state of inflation, a more general (squeezed) state in inflation was considered here. The effect of this squeezed state on the power spectrum was described by the factor $|C(k)|$ (or $|C_g(k)|$), allowing the familiar formulation of the derived power spectrum of curvature perturbations or gravitational waves to be simply multiplied by this factor. Assuming a general initial state in slow-roll inflation, it was shown that the power spectrum can be written in a general form. Using this formulation, particular properties such as oscillation were derived.

The power spectrum was then calculated analytically for finite inflation models with scalar matter- or radiation-dominated pre-inflation. The resultant power spectra for both pre-inflation models exhibit a step-like variation, and in the case of radiation-dominated pre-inflation, the power spectrum oscillates from large to small scales. The ratio of gravitational waves to curvature perturbations in the power spectrum changes from $16\varepsilon$ to $16\varepsilon \,|\,C_g(k)\,|^2 \,/\,|\,C(k)\,|^2$, and is dependent on $z$ (i.e., $k$) at large scales. Calculation of the angular TT and TE power spectra for three typical slow-roll inflation models with scalar matter- or radiation-dominated pre-inflation revealed that the discrepancy between the WMAP3 data and the $\Lambda$ CDM model, that is, suppression of the spectrum at $l = 2$, may be explained to a certain extent by a finite length of inflation in the vicinity of 60 $e$-folds. A small-field inflation model with scalar matter-dominated pre-inflation was found to give the best fit to WMAP3 data among the models considered, and scalar matter-dominated pre-inflation is considered to provide a better result than the radiation-dominated case, independent of the length of inflation. From a general initial state of inflation, the power spectrum can be written according to eqs. (19) and (23), and if the power spectrum oscillates, as in the radiation case, the shape of the TT power spectrum may differ from that given by the $\Lambda$ CDM model. This result implies a constraint on the pre-inflation model.

In both cases of pre-inflation, the angular TE power spectra are similar to those afforded by the $\Lambda$ CDM model. At $l \le 20$, however, the present models produce a lower-amplitude spectrum compared to $\Lambda$ CDM for inflation of close to 60 $e$-folds.

The matching condition employed in the present study for scalar perturbations requires that the gauge potential and its first $\eta$-derivative be continuous at the transition point. Using an alternative matching condition [11] in which the transition is required to occur on a hyper-surface of constant energy, the power spectra exhibit a larger oscillation from large to small scales in both pre-inflation models. These two matching conditions were considered in previous studies of the power spectrum and angular power spectrum under power-law inflation [3,4].

If the length of inflation is finite, and in particular not substantially longer than 60 *e*-folds, the present findings should be taken into account in calculation of the angular power spectrum. In the near future, it is intended that the present analysis will be applied to a more concrete model, such as a super-gravity inspired model [24].

**Acknowledgments**

The authors would like to thank the staff of Osaka Electro-Communication University for valuable discussions, and Mr. Kouta Asajima for use of the program for calculating $\chi^2$.

**Appendix**

Tensor perturbation in the case of radiation-dominated pre-inflation is treated as follows. $|C_g(k)|^2$ can be calculated analytically by

$$|C_g(k)|^2 = A_7 \cosh 2X' + \left(A_8 \cos\left(-2z/q + \theta'\right) + A_9 \sin\left(-2z/q + \theta'\right)\right)\sinh 2X', \quad \text{(A-1)}$$

where

$$A_7 = \frac{\pi}{2z}(2(q-1)zB_1B_2 + z^2B_2^2 + (3 - 2q + q^2 + z^2)B_1^2),$$

$$A_8 = \frac{\pi}{2z}(2(q-1)zB_1B_2 + z^2B_2^2 + (1 - 2q + q^2 - z^2)B_1^2),$$

$$A_9 = \pi B_1((q-1)B_1 + zB_2),$$

$$B_1 = BesselJ\left[1/2 - q, z\right], \ B_2 = BesselJ\left[3/2 - q, z\right], \ z = -k\eta_2, q = -1 - \varepsilon .$$

For $z \to 0$, the leading term is written as

$$|C_g(k)|^2 \approx \frac{\pi 4^{-1+q}(q-1)^2 z^{-2q}}{\Gamma(3/2-q)^2}\{\cosh 2X' + \cos(-2z/q + \theta')\sinh 2X'\} . \tag{A-2}$$

For $z \to \infty$, the leading term is written as

$$|C_g(k)|^2 \approx \cosh 2X' + \cos(q\pi + 2z)\cos(-2z/q + \theta')\sinh 2X' . \tag{A-3}$$

As in the scalar matter-dominated case, $|C_g(k)|^2$ becomes 1 if $X' = 0$.

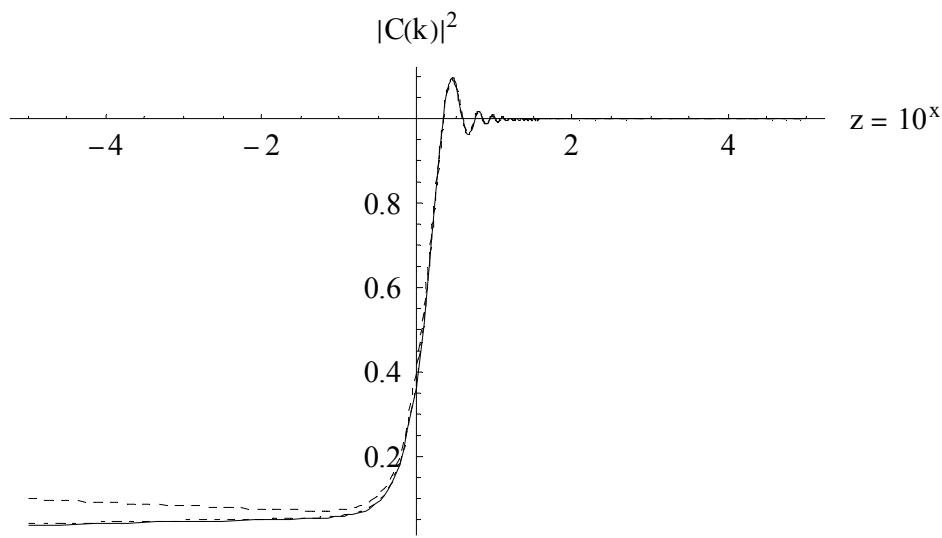

Figure 1. Factor $|C(k)|^2$ as a function of $z = -k\eta_2$ for $10^{-5} \le z \le 10^5$ in the case of a scalar matter-dominated pre-inflation period for three slow-roll inflation models (small-field model, solid line; $\phi^2$ large field model, dash-dotted line; hybrid model, dashed line).

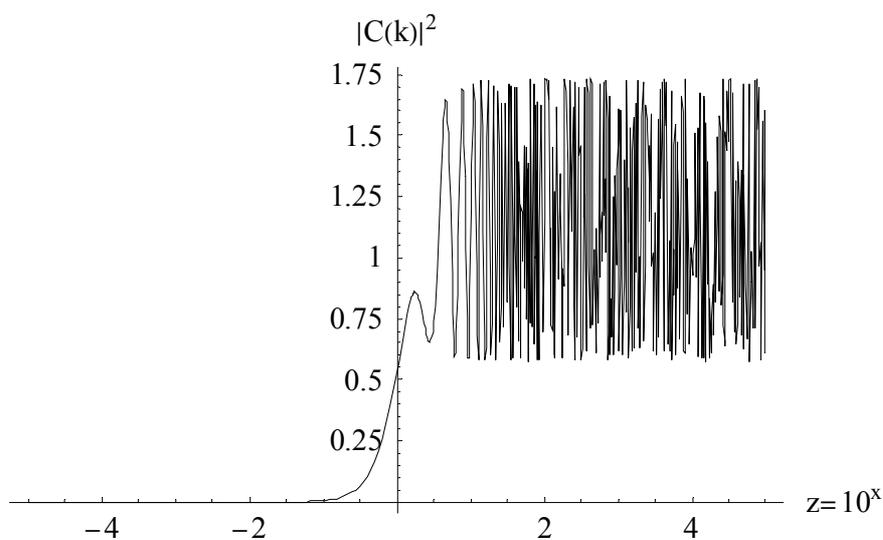

Figure 2. Factor $|C(k)|^2$ as a function of $z = -k\eta_2$ for $10^{-5} \le z \le 10^5$ in the case of a radiation-dominated pre-inflation period (small-field model).

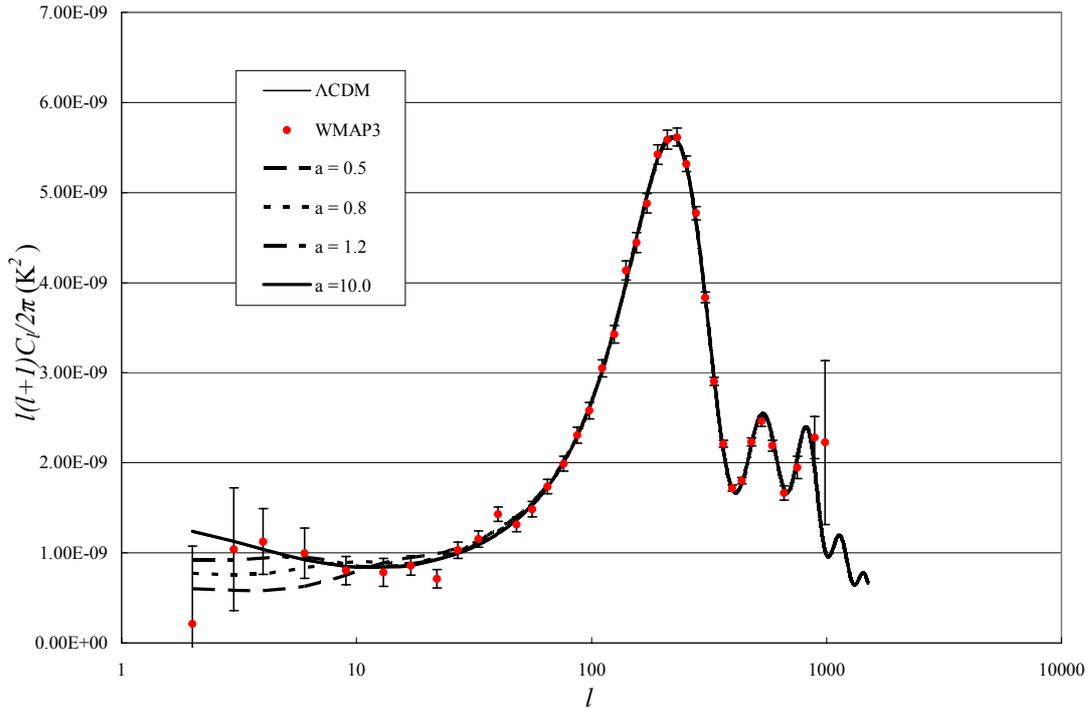

Figure 3. Angular TT power spectrum in the case of a scalar matter-dominated pre-inflation period (small-field model) with $a = 0.5$, $0.8$, $1.2$ and $10.0$. The spectrum afforded by the $\Lambda$CDM model is shown for comparison.

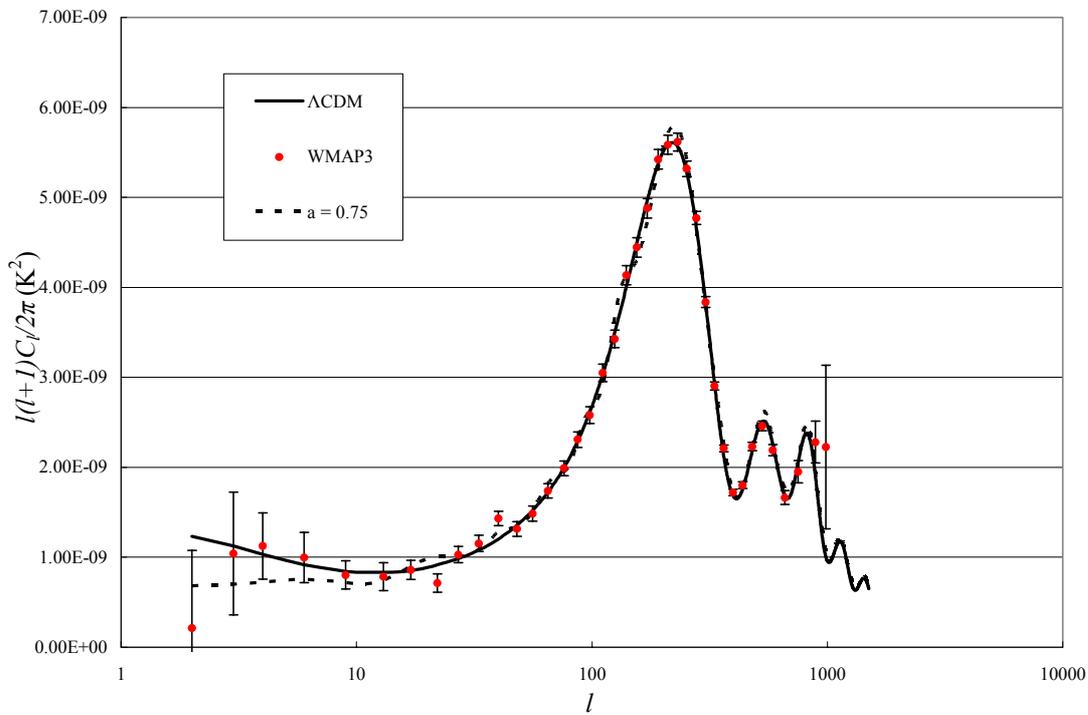

Figure 4. Angular TT power spectrum in the case of a radiation-dominated pre-inflation period with $a = 0.75$. The spectrum afforded by the $\Lambda$ CDM model is shown for comparison.

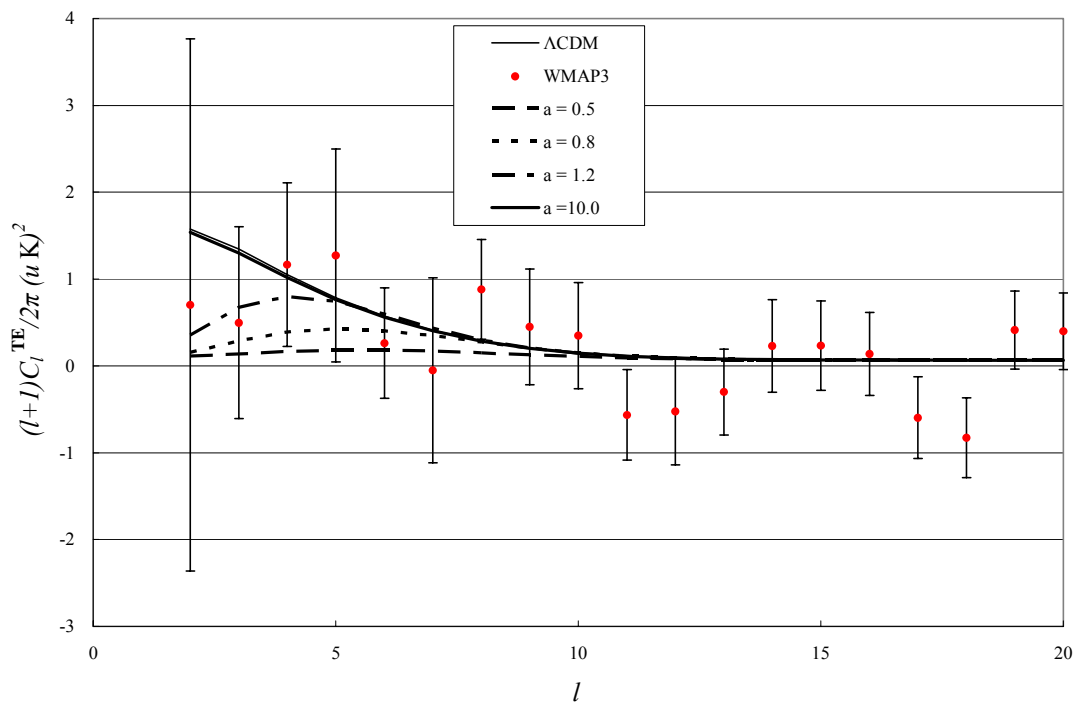

Figure 5. Angular TE power spectrum in the case of a scalar matter-dominated pre-inflation with $a = 0.5$, $0.8$, $1.2$ and $10.0$. The spectrum afforded by the $\Lambda$ CDM model is shown for comparison.